\documentclass[twocolumn,prd,aps,nofootinbib,showpacs]{revtex4}
\usepackage{multirow}
\usepackage{graphicx}
\usepackage[dvips]{color}
\usepackage{color}
\usepackage{amssymb,amsmath}

\begin{document}

\title{Preliminary lattice QCD study of the $I=1$ $K \overline{K}$ scattering length}

\author{Ziwen Fu}
\affiliation{
Key Laboratory of Radiation Physics and Technology {\rm (Sichuan University)}, 
Ministry of Education; \\
Institute of Nuclear Science and Technology; Sichuan University,
Chengdu 610064, P. R. China. 
}

\begin{abstract}
The $s$-wave kaon-antikaon ($K \overline{K}$) elastic scattering length
is investigated by lattice simulation
using pion masses $m_\pi = 330 - 466$ MeV.
Through moving wall sources without gauge fixing,
we calculate $K \overline{K}$ four-point correlation functions
for isospin $I=1$ channel
in the ``Asqtad'' improved staggered fermion formulation,
and observe a clear signal of attraction,
which is consistent with other pioneering lattice studies
on $K \overline{K}$ potential.
Extrapolating $K \overline{K}$ scattering length to the physical point,
we obtain $m_{K} a^{I=1}_{K\overline{K}} = 0.211(33)$.
These simulations are performed with MILC gauge configurations
at lattice spacing $a \approx 0.15$~fm.
\end{abstract}

\maketitle

\section{Introduction}
The elastic $K \overline{K}$ scattering is
one of the simplest reactions with strange quark,
and it allows for an explicit exploration of
the three-flavor  hadronic structure.
The measurement of $K \overline{K}$ scattering
is very useful for our study of the chiral symmetry breaking of
quantum chromodynamics (QCD).
In the isospin limit, the $K \overline{K}$ system
have two isospin eigenstates, namely, $I = 1$ and $0$~\cite{Oller:1998hw}.

During the last decade, the experimental measurements of
the $K \overline{K}$ scattering have been carried out at several cooler-synchrotron COSY facilities~\cite{Winter:2006vd,Maeda:2007cy,
Dzyuba:2008fi,Maeda:2008mx,Silarski:2009yt,Xie:2010md,Lorentz:2011zz}.
Because of the high momentum resolution of COSY, the thresholds from neutral
and charged kaons are well separated and can be studied independently,
and the values of total and differential cross sections
are now available for a variety of reactions~\cite{Winter:2006vd,Maeda:2007cy,
Dzyuba:2008fi,Maeda:2008mx,Silarski:2009yt,Xie:2010md,Lorentz:2011zz},
where the prompt $K^+K^-$ spectrum is described by a four-body
phase-space distribution, modified by the various
final state interaction~\cite{Winter:2006vd,Maeda:2007cy,
Dzyuba:2008fi,Maeda:2008mx,Silarski:2009yt,Xie:2010md,Lorentz:2011zz}.
These results show a clear evidence for the smooth $K^+K^-$ background.

It should be stressed that, based  on the low energy $K^{+}K^{-}$ invariant
mass distributions and generalized Dalitz plot analysis,
Silarski et al. first estimated the scattering length for
the  $K^{+}K^{-}$ interaction
to be $|Re({a_{K^+K^-})}|$ = 0.5$^{+4.0}_{-0.5}$~fm
and $Im({a_{K^+K^-}})$ = 3.0 $\pm$ 3.0~fm~\cite{Silarski:2009yt}.
Later, they improve these results as
$|Re({a_{K^+K^-})}|  = 0.2^{+0.8}_{-0.2}~{\mathrm fm}$,
$Im({a_{K^+K^-}})    = 0.4^{+0.6}_{-0.4}~{\mathrm fm}$
~\cite{Silarski:2010ae}.
Since $K^+K^-$ system is a mixture of the isospin $I=0$ and $1$,
in those analyses we can not distinguish between the $I=0$
and $1$ channels.

The determination of $K \overline{K}$ scattering from QCD is
very difficult since it is essentially a non-perturbative problem.
However, some theoretical efforts are still taken to study
the $K \overline{K}$ scattering~\cite{Wong:1975rd,Guerrero:1998ei,Griffith:1969ph}.
But, if the scattering hadrons contain strange quarks,
the chiral perturbation theory ($\chi$PT) predictions
usually suffer from considerable corrections
because of the resonances $f_0$ and $a_0$,
and one has to apply chiral unitary theory,
sometimes called unitarized chiral perturbation theory~\cite{Oller:1997ti,Oller:1998zr,Doring:2011vk,GomezNicola:2001as}.

The most feasible way to extract $K \overline{K}$ scattering length
nonperturbatively from first principles is using lattice QCD.
And it offers an another important consistent check
of the validity of $\chi$PT with the inclusion of the strange quarks.
Although there are some exploratory lattice QCD investigations
of the meson anti-meson  potential ( including $K \overline{K}$ potential )
in Refs.~\cite{Mihaly:Ph.D,Mihaly:1999ff},
until now, no lattice QCD study about its scattering length
has been reported, mainly because it is extremely difficult to
reliably calculate the rectangular and disconnected diagrams.
Motivated by the first lattice study of the $s$-wave $K^+K^+$
scattering length in the $I=1$ channel,
which was explored by NPLQCD Collaboration in fully-dynamical
lattice QCD with domain-wall valence~\cite{Beane:2007uh}
and the value of $m_{K^+} a_{K^+K^+}$ was found to be $-0.352(16)$,
we will explore the $K \overline{K}$ scattering length from lattice QCD.
This is also encouraged by our reliable extractions of
the $\pi K$~\cite{Fu:2011wc,Fu:2011xw}
and $\pi\pi$~\cite{Fu:2011bz} scattering lengths.

L\"uscher~\cite{Luscher:1991p2480,Lellouch:2001p4241} established
the basic formulae for the calculation of the scattering length
using lattice QCD, which is valid for
the elastic scattering below inelastic thresholds.
Below kinematic thresholds, the scattering length
of two hadrons is connected to the energy phase shift
of two-hadron state enclosed in a torus.
This method paved the way for our lattice calculations
of $K\overline{K}$ scattering.
We should bear in mind that, above the inelastic threshold,
a tower of resonances emerges,
which suggest the opening of other channels.

It is well-known that the appearance of the $f_0(I=0)$ and $a_0(I=1)$,
is due to the introduction
of the $K \overline{K}$ scattering channel~\cite{Oller:1998hw}.
They decay mainly into $\pi\pi$ and $\pi\eta$, respectively.
Both have the masses around $980$~MeV~\cite{Nakamura:2010zzi}.
Since the central mass values fall almost exactly at the $K\overline{K}$ threshold,
the strong coupling to this channel distorts significantly the upper
parts of the mass spectra.

As presented later, for isospin $I = 0$ channel, we need calculating
vacuum diagram, which is extremely difficult to reliably measure.
Thus, we here will preliminarily report our lattice simulations
on the $K \overline{K}$ scattering length in the $I = 1$ channel.
In the presence of the strange quark exchange channel,
the $K \overline{K}$ scattering length have imaginary contribution.
For instance, the experimental determinations of the scattering lengths
are complex as dictated by the various decay channel thresholds~\cite{Winter:2006vd,Maeda:2007cy,
Dzyuba:2008fi,Maeda:2008mx,Silarski:2009yt,Xie:2010md,Lorentz:2011zz}.
The $S$ matrix has the structure
$$
S = \left[
\begin{array}{cc}
\eta e^{2 i \delta_{K \overline{K}}} & i (1 - \eta^2)^{1/2} \,
e^{i (\delta_{K \overline{K}} + \delta_{\pi\eta})} \\
i (1 - \eta^2)^{1/2} \, e^{i (\delta_{K \overline{K}} + \delta_{\pi\eta})} &
\eta e^{2 i \delta_{\pi\eta}}
\end{array}
\right]
$$
where $\delta_{K \overline{K}}$ and $\delta_{\pi\eta}$ are the phase shifts for the
elastic $K \overline{K} \rightarrow K \overline{K}$ and
$\pi\eta \rightarrow \pi\eta$ processes
in isospin $I=1$ channel and $\eta$ is the inelasticity.
We note that, using $(S)_{11}$ and $(S)_{22}$
we can determine $\eta, \delta_{K \overline{K}}$ and $\delta_{\pi\eta}$.

We will use the MILC gauge configurations generated
with the $N_f = 2+1$ flavors of the Asqtad improved~\cite{Orginos:1998ue,Orginos:1999cr}
staggered dynamical sea quarks~\cite{Bernard:2010fr,Bazavov:2009bb}
to calculate the elastic $K \overline{K}$ phase shift $\delta_{K \overline{K}}$
and then evaluate the $s$-wave $K\overline{K}$ scattering length
for isospin $I=1$ channel in the absence of resonant states.
We adopt the technique introduced in Refs.~\cite{Kuramashi:1993ka,Fukugita:1994ve},
namely, the moving wall sources without gauge fixing,
to obtain the reliable accuracy, and observe a clear signal of attraction,
which is consistent with pioneering lattice studies
on $K \overline{K}$ potential in Refs.~\cite{Mihaly:Ph.D,Mihaly:1999ff}.
Moreover, we extrapolate the $K \overline{K}$ scattering
length to the physical point
using  the continuum three-flavor $\chi$PT form
at the next-to-leading order (NLO),
which are provided in ~\ref{sec:AppendixB} with the help of Zhi-Hui Guo,
and it can be directly built from the its scattering amplitudes in Ref.~\cite{Guo:2011pa},

This article is organized as follows.
In Sect.~\ref{sec:Methods}, we describe the formalism for the calculation
of the $K \overline{K}$ scattering lengths
using the L\"uscher's formula~\cite{Luscher:1991p2480,Lellouch:2001p4241}
and our computational technique for $K \overline{K}$ four-point functions.
We present our lattice results in Sect.~\ref{sec:Results},
and arrive at our conclusion and outlook in Sect.~\ref{sec:conclude}.

\section{Method}
\label{sec:Methods}
Here we follow the original notations and conventions
in Refs.~\cite{Nagata:2008wk,Sharpe:1992pp,Kuramashi:1993ka,Fukugita:1994ve,Guo:2011pa}
to review the required formulae
for calculating $K \overline{K}$  scattering lengths.
In the Asqtad-improved staggered dynamical fermion formalism,
let us study the $K \overline{K}$ scattering of
one Goldstone kaon and one Goldstone anti-kaon.
Using operators $O_K(x_1)$, $O_K(x_2)$  for kaons at lattice points $x_1$, $x_2$,
and operators $O_{\overline{K}}(x_3)$, $O_{\overline{K}}(x_4)$
for anti-kaons at lattice points $x_3$, $x_4$, respectively,
we can express the $K \overline{K}$ four-point correlation function  as
\begin{equation}
C_{K \overline{K}}(x_4,x_3,x_2,x_1) =
\bigl< O_{\overline{K}}(x_4) O_{\overline{K}}(x_3)
O_{K}^{\dag}(x_2) O_{K}^{\dag}(x_1)\bigr> ,
\label{EQ:4point_KK}
\end{equation}
where the kaon and anti-kaon interpolating field operators
are denoted by
\begin{eqnarray}
{\cal O}_{K^+}(t) &=& \sum_{\bf{x}} \bar{s}({\bf{x}},t)\gamma_5 u({\bf{x}},t) ,\cr
{\cal O}_{K^0}(t) &=& \sum_{\bf{x}} \bar{s}({\bf{x}},t)\gamma_5 d({\bf{x}},t) , \cr
{\cal O}_{\overline{K}^0}(t) &=&
-\sum_{\bf{x}} \bar{d}({\bf{x}},t)\gamma_5 s({\bf{x}},t) ,\cr
{\cal O}_{K^-}(t) &=& \sum_{\bf{x}} \bar{u}({\bf{x}},t)\gamma_5 s({\bf{x}}, t) .
\label{EQ:IFO_K}
\end{eqnarray}
After summing over the spatial coordinates,
we achieve the $K \overline{K}$ four-point correlation function,
\begin{equation}
\label{EQ:4point_KK_k000}
C_{K \overline{K}}(t_4,t_3,t_2,t_1) =
\sum_{{\bf{x}}_1} \sum_{{\bf{x}}_2}\sum_{{\bf{x}}_3}\sum_{{\bf{x}}_4}
C_{K \overline{K}}(x_4,x_3,x_2,x_1) ,
\end{equation}
where $x_1 \equiv ({\bf{x}}_1,t_1)$ (likewise for $x_2$, $x_3$, and $x_4$),
and $t$ represents the time difference, namely, $t\equiv t_3 - t_1$.
We build $K \overline{K}$ operators for
two isospin eigenstates as~\cite{Oller:1998hw}~\footnote{
Our phase conventions for pseudoscalar mesons are different
from those in Ref.~\cite{Oller:1998hw}.
}
\begin{eqnarray}
\left.|K \overline{K}(I=0)\right\rangle  = \sqrt{\frac{1}{2}}\left\{
\left.|K^+K^-\right\rangle - \left. | K^0 \overline{K^0} \right\rangle \right\}, \cr
\left.|K \overline{K}(I=1)\right\rangle = \sqrt{\frac{1}{2}}\left\{
\left.|K^+K^-\right\rangle + \left.| K^0 \overline{K^0} \right\rangle \right\} . \nonumber
\end{eqnarray}
Then, the scattering amplitudes $M$ for $K \overline{K}$ scattering
in the $I=1$ and $0$ states are given by
\begin{eqnarray}
M(I\hspace{-0.15cm}=\hspace{-0.15cm}1) \hspace{-0.2cm}&=&\hspace{-0.2cm}
\frac{1}{2} \langle K^+K^-|S|K^+K^-\rangle +
\frac{1}{2} \langle K^+K^-|S|K^0 \overline{K^0}\rangle \cr
&+&\hspace{-0.2cm}
\frac{1}{2} \langle K^0 \overline{K^0}|S|K^+K^-\rangle  +
\frac{1}{2} \langle K^0 \overline{K^0}|S|K^0 \overline{K^0}\rangle ,
\label{EQ:4point_KK_M_I1} \\
M(I\hspace{-0.15cm}=\hspace{-0.15cm}0) \hspace{-0.2cm}&=&\hspace{-0.2cm}
\frac{1}{2} \langle K^+K^-|S|K^+K^-\rangle -
\frac{1}{2} \langle K^+K^-|S|K^0 \overline{K^0}\rangle \cr
&-&\hspace{-0.2cm}
\frac{1}{2} \langle K^0 \overline{K^0}|S|K^+K^-\rangle  +
\frac{1}{2} \langle K^0 \overline{K^0}|S|K^0 \overline{K^0}\rangle .
\label{EQ:4point_KK_M_I0}
\end{eqnarray}

\begin{figure}[thb]
\includegraphics[width=8cm,clip]{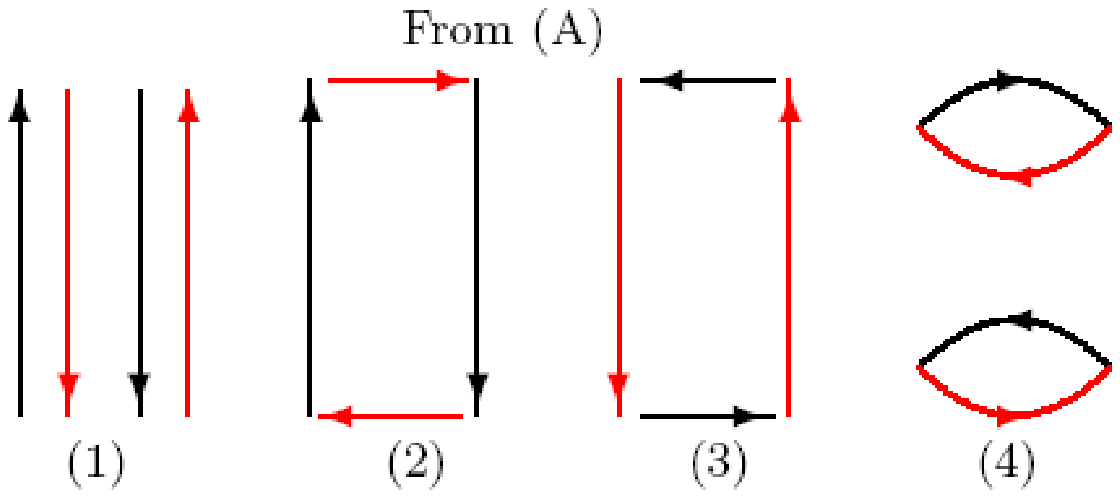}
\includegraphics[width=8cm,clip]{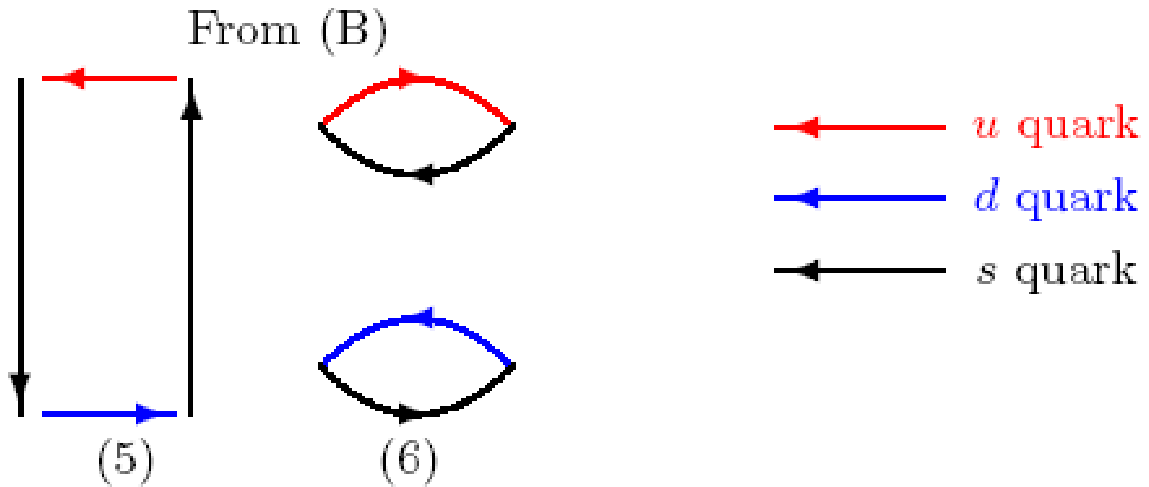}
\caption{ \label{fig:diagram_AB}
The numbers below the quark line diagrams correspond to the terms
(\ref{appA:term_01}) - (\ref{appA:term_06}) given in Appendix A.
(A) and (B) correspond to the first and second terms
in Eq.~(\ref{EQ:4point_KK_M_I1}), respectively.
}
\end{figure}

\begin{figure}[thb]
\includegraphics[width=8cm,clip]{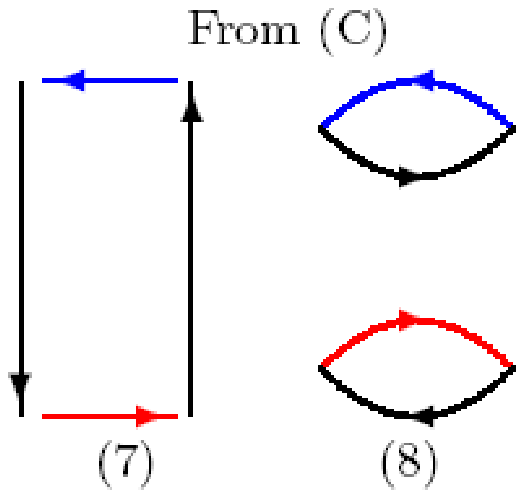}
\includegraphics[width=8cm,clip]{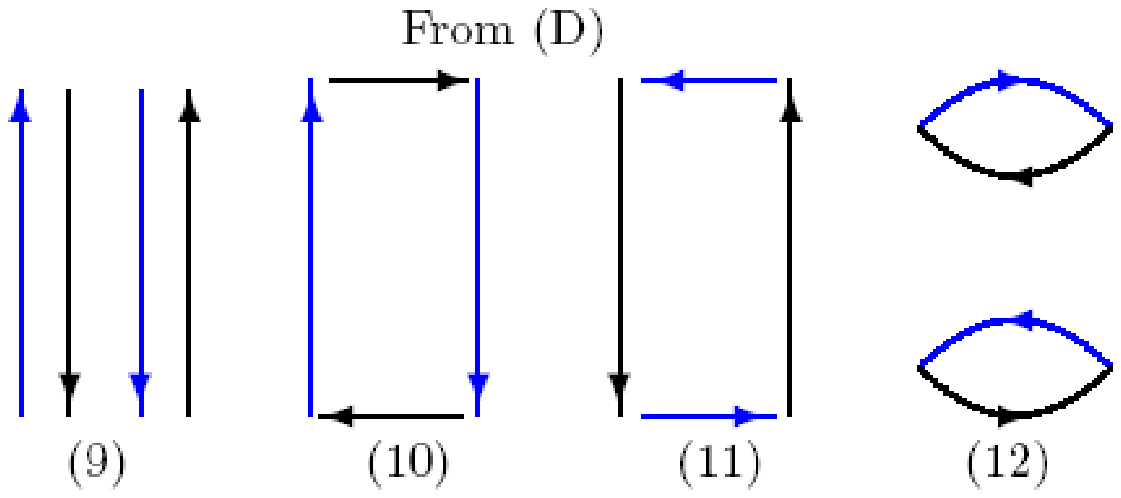}
\caption{ \label{fig:diagram_CD}
The numbers below the quark line diagrams correspond to the terms
(\ref{appA:term_07}) - (\ref{appA:term_12}) given in Appendix A.
(C) and (D) correspond to the third and fourth terms
in Eq.~(\ref{EQ:4point_KK_M_I1}), respectively.
}
\end{figure}

We substitute Eq.~(\ref{EQ:IFO_K}) into Eq.~(\ref{EQ:4point_KK}) to obtain
the quark diagrams for the $K \overline{K}$ scattering in the $I=1$ channel.
The concrete calculations are given in~\ref{sec:AppendixA} for reference.
At last, we obtain $12$ different diagrams for the $I=1$ channel,
which correspond to Eqs.~(\ref{appA:term_01}) - (\ref{appA:term_12}),
and are also shown in Figs.~\ref{fig:diagram_AB} and \ref{fig:diagram_CD},

In the isospin limit,
we can classify the $12$ quark line diagrams into four independent groups.
Diagram $1$ in Fig.~\ref{fig:diagram_AB} and diagram $9$
in Fig.~\ref{fig:diagram_CD} are categorized into Group $1$.
Likewise, diagrams $3$ and $10$ are categorized into Group 2;
Nos.\ $2, 5, 7$, and $11$ in Group $3$;
Nos.\ $4, 6, 8$, and $12$ in Group $4$; respectively.
By inspecting Eq.~(\ref{EQ:4point_KK_M_I1}) and
Eqs.~(\ref{appA:term_01}) - (\ref{appA:term_12}),
the coefficients of these groups are given as
\begin{eqnarray}
\hbox{Group 1:} &&\cr
&&\frac{1}{2} + \frac{1}{2} = 1 : D, \cr
\hbox{Group 2:} &&\cr
&& \frac{1}{2}(-1) + \frac{1}{2}(-1) = -1 : R_u,  \cr
\hbox{Group 3:} &&\\
&& \frac{1}{2} + \left(\frac{1}{2}\right)(-1) + \left(\frac{1}{2}\right)(-1)
+ \frac{1}{2} = 0 : R_s, \cr
\hbox{Group 4:} &&\cr
&& \frac{1}{2} + \left(\frac{1}{2}\right)(-1) + \left(\frac{1}{2}\right)(-1)
+ \frac{1}{2} = 0 : V, \nonumber
\end{eqnarray}
where,  $R_u$, $R_s$ are two types of the rectangular quark diagrams
and the subscript $u$ and $s$ distinguish two of them,
since for $R_u$ rectangular quark diagram,
there are two $u$ quark lines on the time direction, likewise for $R_s$.
At last, only two diagrams still contribute to
$K \overline{K}$ scattering amplitudes for the isospin $I=1$ channel.
For the $I=0$ case, we can easily perform the same procedures.
Finally, both channels can be written
in terms of the diagrams $D$, $R_u$, $R_s$ and $V$,
namely~\cite{Mihaly:Ph.D,Mihaly:1999ff}
\begin{eqnarray}
M(I=1) &=& D - R_u ,   \cr
M(I=0) &=& D - R_u - 2 R_s + 2 V .
\end{eqnarray}

To avert the Fierz rearrangement of quark lines~\cite{Kuramashi:1993ka,Fukugita:1994ve},
we select $t_1 =0$, $t_2=1$, $t_3=t$, and $t_4 = t+1$.
Then, we can build $K \overline{K}$ operators for
isospin $I = 1$ and $0$ eigenstates as~\cite{Oller:1998hw}
\begin{eqnarray}
\label{EQ:op_pipi}
{\cal O}_{K \overline{K} }^{I=0} (t)\hspace{-0.2cm} &=&\hspace{-0.2cm}\frac{1}{\sqrt{2}}
\left\{ K^+(t) K^-(t+1) \hspace{-0.1cm}-\hspace{-0.1cm} K^0(t) \overline{K}^0(t+1)  \right\} , \cr
{\cal O}_{K \overline{K} }^{I=1} (t)
\hspace{-0.2cm}&=&\hspace{-0.2cm}\frac{1}{\sqrt{2}}
\left\{ K^+(t) K^-(t+1) \hspace{-0.1cm}+\hspace{-0.1cm} K^0(t) \overline{K}^0(t+1)  \right\} ,
\end{eqnarray}

\begin{figure}[thb]
\includegraphics[width=8cm,clip]{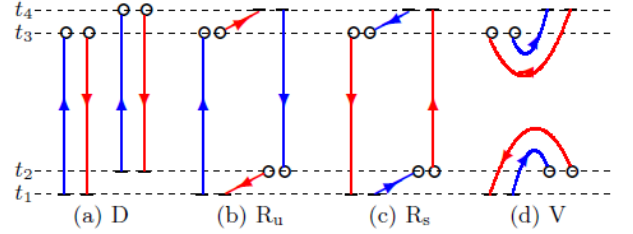}
\caption{ \label{fig:diagram_KK}
Diagrams contributing to $K \overline{K}$ four-point functions.
Short bars stand for wall sources.
Open circles are sinks for local kaon or anti-kaon operators.
The blue and red lines represent the $u/d$ and strange quark lines, respectively.
}
\end{figure}
In the isospin limit, four quark line diagrams contribute to
the $K \overline{K}$ scattering amplitudes;
we plot them in Fig.~\ref{fig:diagram_KK},
and label them as direct ($D$), rectangular ($R_u$),
rectangular ($R_s$), and vacuum ($V$) diagrams, respectively.
It is pretty easy to evaluate the direct diagram,
while the reliable evaluation of the rectangular ($R_u, R_s$) and
vacuum ($V$) diagrams is extremely difficult~\cite{Kuramashi:1993ka,Fukugita:1994ve}.
We settle it through the moving wall source technique
introduced in Refs.~\cite{Kuramashi:1993ka,Fukugita:1994ve}, namely,
each propagator, which corresponds to a moving wall source
at $ t = 0, \cdots, T-1$, is denoted by
$$
\sum_{n''}D_{n',n''}G_t(n'') = \sum_{\bf{x}}
\delta_{n',({\bf{x}},t)}, \quad 0 \leq t \leq T-1 ,
$$
where $D$ defines the quark matrix for the quark action.
$D$, $R_u$, $R_s$ and $V$ are schematically displayed in Fig.~\ref{fig:diagram_KK},
and we can also write them in terms of the quark propagators $G$, namely,
\begin{eqnarray}
\label{eq:dcr}
C_D(t_4,t_3,t_2,t_1) &=&  \sum_{ {\bf{x}}_3}\sum_{ {\bf{x}}_4 }
\langle \mbox{Tr}
[G_{t_1}^{(u)\dag}({\bf{x}}_3, t_3) G^{(s)}_{t_1}({\bf{x}}_3, t_3)] \cr
&&\times
\mbox{Tr}[G_{t_2}^{(u)\dag}({\bf{x}}_4, t_4) G^{(s)}_{t_2}({\bf{x}}_4, t_4) ] \rangle,\cr
C_{R_u}(t_4,t_3,t_2,t_1) &=&  \sum_{ {\bf{x}}_2}\sum_{ {\bf{x}}_3 }
\langle \mbox{Tr}
[G_{t_1}^{(s)\dag}({\bf{x}}_2, t_2) G^{(u)}_{t_4}({\bf{x}}_2, t_2)  \cr
&&\times
G_{t_4}^{(s)\dag}({\bf{x}}_3, t_3) G^{(u)}_{t_1}({\bf{x}}_3, t_3) ] \rangle, \cr
C_{R_s}(t_4,t_3,t_2,t_1) &=&  \sum_{ {\bf{x}}_2}\sum_{ {\bf{x}}_3 }
\langle \mbox{Tr}
[G_{t_1}^{(u)\dag}({\bf{x}}_2, t_2) G^{(s)}_{t_4}({\bf{x}}_2, t_2)  \cr
&&\times
G_{t_4}^{(u)\dag}({\bf{x}}_3, t_3) G^{(s)}_{t_1}({\bf{x}}_3, t_3) ] \rangle, \cr
C_V(t_4,t_3,t_2,t_1) &=& \sum_{{\bf{x}}_2} \sum_{ {\bf{x}}_3}
\Bigl\{ \langle \mbox{Tr}
[G_{t_1}^{(u) \dag}({\bf{x}}_2, t_2) G_{t_1}^{(s)}({\bf{x}}_2, t_2) ] \cr
&& \times \mbox{Tr}
[G_{t_4}^{(u) \dag}({\bf{x}}_3, t_3) G_{t_4}^{(s)}({\bf{x}}_3, t_3) ] \rangle  -\cr
&&\langle \mbox{Tr}
[G_{t_1}^{(u) \dag}({\bf{x}}_2, t_2)
 G_{t_1}^{(s)}({\bf{x}}_2, t_2)] \rangle \cr
&&\times \langle \mbox{Tr}
[G_{t_4}^{(u) \dag} ({\bf{x}}_3, t_3) G_{t_4}^{(s)}({\bf{x}}_3, t_3) ]
\rangle \Bigl\} ,
\end{eqnarray}
where we utilize the hermiticity properties of $G$
to remove the $\gamma^5$ factors.
The vacuum diagram here include a vacuum subtraction term.

According to the discussions
in Refs.~\cite{Kuramashi:1993ka,Fukugita:1994ve},
the $K \overline{K}$ rectangular and vacuum diagrams
in Fig.~\ref{fig:diagram_KK}
produce the gauge-variant noise,
and we will perform the gauge field average without gauge fixing
to nicely suppress this noise.
We can construct physical $K \overline{K}$ correlators with certain isospin
by integrating four types of the propagators.
In the isospin limit, we can express the $K \overline{K}$ correlator
for isospin $I=1$ and $0$ channels by
\begin{eqnarray}
\label{EQ:phy_I12_32}
C_{K \overline{K}}^{I=0}(t) &\equiv&
\left\langle {\cal O}_{K \overline{K}}^{I=0} (t) | {\cal O}_{K \overline{K}}^{I=0} (0) \right\rangle
= D - R_u - 2R_s + 2V, \cr
C_{K \overline{K} }^{I=1}(t) &\equiv&
\left\langle {\cal O}_{K \overline{K}}^{I=1} (t) | {\cal O}_{K \overline{K}}^{I=1} (0) \right\rangle
= D - R_u ,
\end{eqnarray}
where operator ${\cal O}_{K \overline{K}}^{I}$
creates a $K \overline{K}$ state with total isospin $I$.
Since, in this paper, we preliminarily report
our lattice simulations on $K \overline{K}$ scattering length
in the $I = 1$ channel,
in the following discussions, we will remove the superscript $I$
for the corresponding quantities.

There are some different methods to parameterize the low-momentum
behavior of the scattering amplitude.
To calculate the $K \overline{K}$ scattering lengths on the lattice,
we adopt the L\"uscher formula,
and  it is straightforward to use the standard effective range expansion
for the $K \overline{K}$ scattering phase shift, namely,
\begin{equation}
\label{eq:exact}
k  \cot \delta_0(k) = \frac{1}{a} + \frac{1}{2} r k^2 + {\cal O}(k^4) ,
\end{equation}
where $\delta_0(k)$ is $s$-wave scattering phase shift,
$a$ is the scattering length,
$r$ is the effective range,
and $k$ is the magnitude of the center-of-mass scattering momentum
related to the total energy of the $K \overline{K}$ system
in a cubic box of size $L$ by
$E_{K \overline{K}} = 2\sqrt{m_K^2 + k^2}$,
where $m_K$ is kaon mass. This expansion is only valid at low energy,
unfortunately, takes inelasticities not into consideration.

If we approximate the $K \overline{K}$  element of
the S matrix by its dominant pole contribution,
the $s$-wave $K \overline{K}$ scattering length in the continuum
is denoted by employing the standard L\"uscher formula
$$
a_0 = \lim_{k\to 0} \frac{\tan\delta_0(k)}{k} ,
$$
where the scattering is purely elastic below inelastic thresholds.
We should bear in mind that the truncation of the effective range $r$
in Eq.~(\ref{eq:exact}) serves as a source of the systematic uncertainty,
which appears as ${\cal O}(1/L^6)$ and in this paper we ignore it.
$\delta_0(k)$ is the $s$-wave scattering phase shift,
which can be evaluated by the L\"uscher's finite size
formula~\cite{Luscher:1991p2480,Lellouch:2001p4241},
\begin{equation}
\left( \frac{\tan\delta_0(k)}{k} \right)^{-1} =
\frac{\sqrt{4\pi}}{\pi L} \cdot {\mathcal Z}_{00}
\left(1,\frac{k^2}{(2\pi/L)^2}\right) \,,
\label{eq:luscher}
\end{equation}
where the zeta function $\mathcal{Z}_{00}(1;q^2)$ is denoted by
\begin{equation}
\label{eq:Z00d}
\mathcal{Z}_{00}(1;q^2) = \frac{1}{\sqrt{4\pi}}\sum_{{\mathbf n}\in\mathbb{Z}^3} \frac{1}{n^2-q^2} ,
\end{equation}
here $q = kL/(2\pi)$, and in this work
the zeta function $\mathcal{Z}_{00}(1;q^2)$  is calculated
by the method in Ref.~\cite{Yamazaki:2004qb}.

The energy $E_{K \overline{K}}$ of the $K \overline{K}$ system
can be obtained from the $K \overline{K}$ four-point correlator.
At large  $t$ these correlators
behave as~\cite{Barkai:1985gy,Mihaly:1996ue}
\begin{eqnarray}
\label{eq:E_KaK}
C_{K \overline{K}}(t)  &=&
Z_{K \overline{K}}\cosh\left[E_{K \overline{K}}
\left(t - \frac{T}{2}\right)\right] +\cr
&&
(-1)^t Z_{K \overline{K}}^{\prime}\cosh\left[E_{K \overline{K}}^{\prime} \left(t-\frac{T}{2}\right)\right] + \cdots.
\end{eqnarray}
where $E_{K \overline{K}}$ is the energy of
the lightest $K \overline{K}$ state.
The second term alternating in sign is a peculiarity of
the staggered scheme~\cite{Barkai:1985gy,Mihaly:1996ue}.

In our concrete calculation we also measure the energy shift
$\delta E = E_{K \overline{K}} - 2m_K$ from the ratio
\begin{equation}
\label{EQ:ratio}
R^X(t) = \frac{ C_{K \overline{K}}^X(0,1,t,t+1) }
{ C_K (0,t) C_K(1,t+1) },
\quad  X\hspace{-0.05cm}=D \ {\rm and} \ R_u ,
\end{equation}
where $C_K $ is the kaon two-point correlator.
Considering equation~(\ref{EQ:phy_I12_32}),
we can write the amplitudes which project out the $I=1$
isospin eigenstate as
\begin{equation}
\label{EQ:proj_I0I2}
R(t) = R^D(t) - R^{R_u}(t) .
\end{equation}

We should stress that,
as compared to the contributions of the Nambu-Goldstone kaon and anti-kaon,
these of the non-Nambu-Goldstone kaons and anti-kaons
in the intermediate states are exponentially suppressed
for large times due to their heavier masses~\cite{Sharpe:1992pp,Kuramashi:1993ka,Fukugita:1994ve}.
In the current study we assume that the $K \overline{K}$ interpolator
does not couple significantly to other $K \overline{K}$ tastes,
and ignore this systematic error.

\section{Simulation results}
\label{sec:Results}
We calculated $K \overline{K}$ correlator
on MILC lattice ensemble of $200$ $16^3 \times 48$ gauge configurations
with bare gauge coupling $10/g^2 = 6.572$
and bare quark masses $am_{ud}/am_s = 0.0097/0.0484$.
Its physical volume is about $2.5$ fm.
The inverse lattice spacing $a^{-1} = 1.358^{+35}_{-13}$  GeV~\cite{Bernard:2010fr,Bazavov:2009bb} ( about $0.15$~fm ).
The dynamical strange quark mass is quite close to its physical
value~\cite{Bernard:2010fr,Bazavov:2009bb}.

The necessary matrix element of inverse fermion matrix are
computed using the standard conjugate gradient method.
We calculate the $K \overline{K}$ correlators on
all the time slices  for both source and sink.
After averaging over all its possible values,
the statistics are dramatically enhanced
because we can place the kaon and anti-kaon sources at all time slices.

Using the same lattice gauge configurations,
we calculate the $K \overline{K}$ four-point correlation functions
with six light $u$ valence quarks, namely,
$am_x = 0.0097$, $0.01067$, $0.01261$, $0.01358$, $0.01455$ and $0.0194$,
where we adopt MILC convention: $m_x$ is the valence $u/d$ quark mass.
The strange sea quark mass is chosen at its
physical number~\cite{Bazavov:2009bb}.

\begin{figure}[h]
\includegraphics[width=8cm,clip]{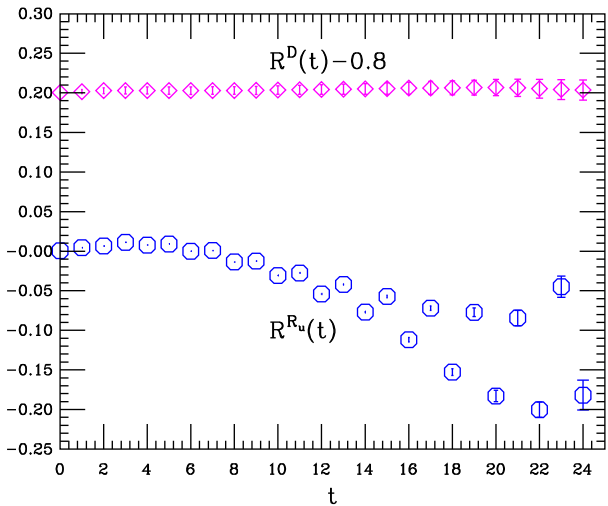}
\caption{
Individual amplitude ratios $R^X(t)$ for $K \overline{K}$
four-point function measured by the moving wall source
without gauge fixing as functions of $t$:
Direct diagram shifted by $0.8$ (diamonds)
and rectangular (octagons) diagrams.
\label{fig:ratio}
}
\end{figure}

In Fig.~\ref{fig:ratio} the individual ratios, $R^X$ ($X=D$ and $R_u$)
are shown as the functions of $t$ for $am_x=0.0097$.
The values of the direct amplitude $R^D$ is quite close to unity,
suggesting the weak interaction in this channel.
After a beginning increase up to $t \sim 4$, the rectangular amplitude
shows an approximately linear decrease up until $t \sim 20$,
indicating an attractive force between the kaon and anti-kaon in this channel.
These features are what we expected from
the theoretical predictions~\cite{Sharpe:1992pp}.
The systematically oscillating characteristics
in the large time region is also distinctly noticed,
which is a speciality of the staggered
scheme~\cite{Barkai:1985gy,Mihaly:1996ue}.

In our previous works~\cite{fzw:2011cpl,fzw:2011cpc12,Fu:2011wc},
we have measured the pion and kaon point-to-point correlators,
and calculated the pion mass $m_\pi$, kaon mass $m_K$,
and pion decay constants $f_\pi$,
which are summarized in Table I of Ref.~\cite{Fu:2011wc}.
In this work, we will directly quote these values.

\begin{figure}[h!]
\includegraphics[width=8cm,clip]{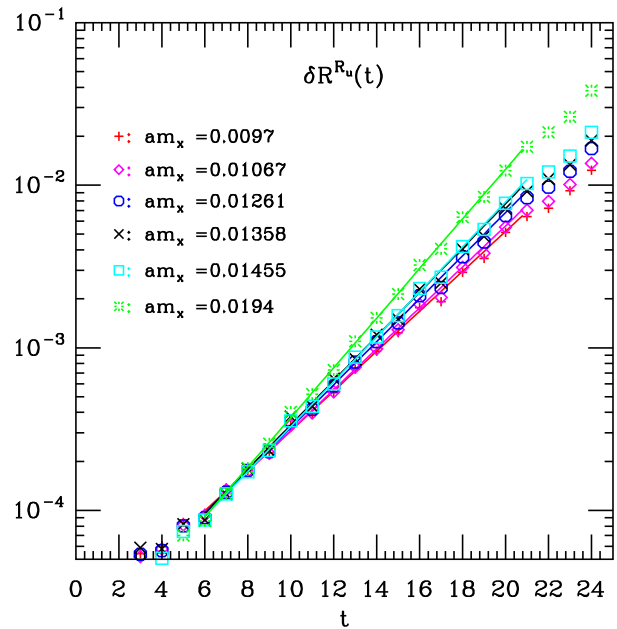}
\vspace{-0.3cm}
\caption{\label{fig:error_R}
The errors of the ratios $R^{R_u}(t)$ as the functions of $t$.
Solid lines are single exponential fits.
}
\end{figure}
According to the arguments in Ref.~\cite{DeGrand:1990ss},
the ratios for the rectangular $R_u$ diagram have errors,
which increase exponentially as $\displaystyle e^{m_\pi t}$
from large time separation~\footnote{
The errors for vacuum amplitudes should be roughly independent of $t$,
and grows exponentially as $\displaystyle e^{2m_K t}$
in the ratio $R_{V}$.
While for the rectangular diagram $R_s$,
it increases as $\displaystyle e^{m_{s\bar{s}} t}$,
where $m_{s\bar{s}}$ is a fictitious meson with
two valence quarks with mass about $m_s$.
Thus, the reliable calculations of these terms are
beyond the scope of this paper
since it requires a substantial amount of computing resources.
}.
The magnitude of the errors is in quantitatively agreement with
this expectation as displayed in Fig.~\ref{fig:error_R}.
Fitting the errors $\delta  R^{R_u}(t)$
by a single exponential fit ansatz
$\displaystyle \delta R^{R_u}(t) \sim  e^{\mu_R t}$,
we can obtain the corresponding fitting values of $\mu_R$.
The fitted values of $\mu_R$ in lattice units and its fitting ranges
are summarized in Table~\ref{tab:error_fitting}.
From Table~\ref{tab:error_fitting}, we can note that the fitted values of
$\mu_R$  can be compared with the corresponding pion masses $m_\pi$
listed in Table I of Ref.~\cite{Fu:2011wc}.
This demonstrates, on the other side,
that the technique of the moving wall source without gauge fixing
used in the current work for the $K \overline{K}$ scattering
is practically feasible.
\begin{table}[h]	
\caption{\label{tab:error_fitting}
Summary of the fitted values of $\mu_R$ in lattice units.
The second block shows the fitted values for $\mu_R$,
Column three shows the time range for the chosen fit.
}
\begin{tabular*}{\columnwidth}{@{\extracolsep{\fill}}lll@{}}
\hline
$am_x$     & $a\mu_R$ & Range  \\
\hline
$0.0097$   & $0.2768$  & $8-14$  \\
$0.01067$  & $0.2825$  & $8-14$  \\
$0.01261$  & $0.3047$  & $8-14$  \\
$0.01358$  & $0.3134$  & $8-14$  \\
$0.01455$  & $0.3183$  & $8-14$  \\
$0.0194$   & $0.3518$  & $8-14$  \\
\hline
\end{tabular*}
\end{table}

\begin{figure}[thb!]
\includegraphics[width=8cm,clip]{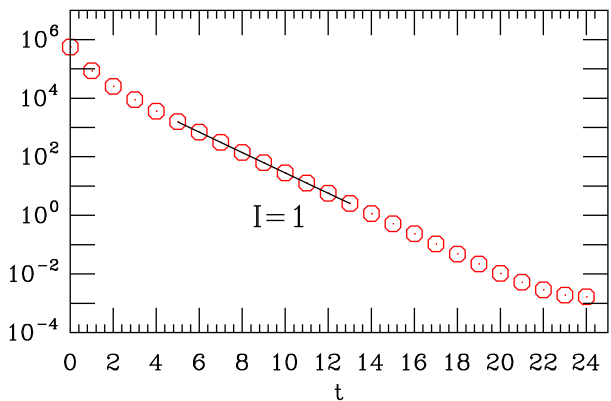}
\caption{
The $K \overline{K}$ correlator
calculated with the moving wall source without gauge fixing
for $am_x=0.0097$.
Solid line is the fit for $6 \le t \le 12$
using the fitting model in Eq.~(\ref{eq:E_KaK}).
\label{fig:I0I2}
}
\end{figure}

As we practiced in Ref.~\cite{Fu:2011wc}, we understood that
the correctly extracting these energies is very important to
our ultimate results, and in our concrete simulation,
they were chosen by searching for the combination of a ``plateau'' in the energy
as the function of the minimum distance,
and a good chi-square (namely, $\chi^2$).
In Fig.~\ref{fig:I0I2} we plot the $K \overline{K}$ correlation
function for $am_x=0.0097$,
where we can compare the fitted functional form
with the lattice data.
The fitted values of the energies $a E$, fitting range and fitting quality
are listed in Table~\ref{tab:fitting_results_I1}.

\begin{table}[h]	
\caption{\label{tab:fitting_results_I1}
Summary of the lattice results for the fitted values of the energy.
The third block shows its fit range, and the fourth block
gives its fit quality $\chi^2/\mathrm{dof}$.
}
\begin{tabular*}{\columnwidth}{@{\extracolsep{\fill}}llll@{}}
\hline
$a m_x$   & $a E$ &  $\mathrm{Range}$  & $\chi^2/{\rm dof}$ \\
\hline
$0.00970$  & $0.7862(6)$  & $6-12$  & $4.29/3$  \\
$0.01067$  & $0.7931(6)$  & $6-12$  & $4.81/3$  \\
$0.01261$  & $0.8071(6)$  & $6-12$  & $4.96/3$  \\
$0.01358$  & $0.8139(6)$  & $6-12$  & $5.01/3$  \\
$0.01455$  & $0.8207(6)$  & $6-12$  & $5.12/3$  \\
$0.01940$  & $0.8538(5)$  & $6-12$  & $5.58/3$  \\
\hline
\end{tabular*}
\end{table}

We now can plug these energies in Table~\ref{tab:fitting_results_I1}
into Eq.~(\ref{eq:exact}) to achieve the scattering lengths.
The center-of-mass scattering momentum $k^2$ in GeV
and the scattering lengths
are tabulated in Table~\ref{tab:I1_2}.
Here we utilize the kaon masses given in Table I of Ref.~\cite{Fu:2011wc}.
The errors of the scattering momentum $k$ and the scattering lengths
are estimated from the statistic errors of
the energies $E$ and kaon masses $m_K$.

\begin{table}[h]	
\caption{\label{tab:I1_2}
Summary of lattice results for the scattering lengths.
Column two gives the center-of-mass scattering momentum $k^2$ in GeV,
and the third block shows the kaon mass times the scattering lengths.
}
\begin{tabular*}{\columnwidth}{@{\extracolsep{\fill}}lll@{}}
\hline
$am_x$   &  $k^2$[${\rm GeV}^2$]  & $m_K a_{K \overline{K}}$\\
\hline
$0.00970$  &  $-0.00444(54)$  & $0.363(52)$  \\
$0.01067$  &  $-0.00446(53)$  & $0.368(52)$  \\
$0.01261$  &  $-0.00451(52)$  & $0.378(52)$  \\
$0.01358$  &  $-0.00483(53)$  & $0.414(54)$  \\
$0.01455$  &  $-0.00471(54)$  & $0.406(55)$  \\
$0.01940$  &  $-0.00488(54)$  & $0.441(58)$ \\
\hline
\end{tabular*}
\end{table}

\begin{figure}[h!]
\includegraphics[width=80mm]{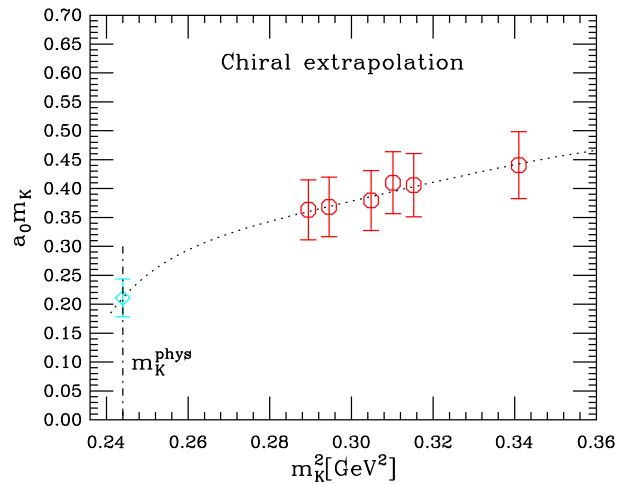}
\caption{\label{fig:ChPT-fit}
$m_K^2$-dependence of the $K \overline{K}$ scattering lengths
$m_K a_{K \overline{K}}$ for the $I=1$ channel.
The cyan diamond point indicate its physical values.
}
\end{figure}

In Fig.~\ref{fig:ChPT-fit}, the $s$-wave $K \overline{K}$ scattering lengths
$m_K a_{K \overline{K}}^{I=1}$
are displayed as a function of $m_K^2$.
In this work, we use pion masses  $m_\pi = 330 - 466$ MeV,
and need to extrapolate the $K \overline{K}$
scattering lengths toward the physical point.
For this end, in ~\ref{sec:AppendixB} we provide the continuum
$\rm SU(3)$ $\chi$PT form at NLO for $a_{K \overline{K}}^{I=1}$,
which can be directly built from the its scattering amplitudes in Ref.~\cite{Guo:2011pa}, namely,
\begin{eqnarray}
\label{eq:mKa1_alg}
m_K a_{K \overline{K}}^{I=1} &=& \frac{m_K^2}{8\pi f_\pi^2}
\bigg\{ 1 + \frac{16 m_\pi^2}{f_\pi^2}L_5(\mu) +
\frac{32 m_K^2}{f_\pi^2} L_{K \overline{K}}^{I=1}(\mu) \cr
&&+\frac{1}{16\pi^2 f_\pi^2} \chi_{K \overline{K}}^{I=1}(\mu)
\bigg\} ,
\end{eqnarray}
where we substituted the pion mass $m_\pi$,
kaon mass $m_K$ and pion decay constant $f_\pi$
listed in Table I of Ref.~\cite{Fu:2011wc}.
$L_5(\mu)$ and $L_{K \overline{K}}^{I=1}(\mu) \equiv 2L_1 + 2L_2 + L_3 - 2L_4
 -\frac{1}{2}L_5 + 2L_6 + L_8$ are
low-energy constants denoted in Ref.~\cite{Gasser:1984gg}
and explicitly dependent on the chiral scale $\mu$.
The $\chi_{K \overline{K}}^{I=1}(\mu)$
is the known functions at NLO including chiral logarithm terms,
see Eq.~(\ref{eq:chi_kk_1}) for details.

To enhance the $\chi$PT fit,
we include all the lattice simulation data
of the $K \overline{K}$ scattering lengths.
The fitting results of the $K \overline{K}$ scattering lengths
$m_K a_{K \overline{K}}^{I=1}$ are plotted by the dotted lines
as a function of $m_K^2$ in Fig.~\ref{fig:ChPT-fit}.
The chirally extrapolated  $K \overline{K}$ scattering length
$m_K a_{K \overline{K}}^{I=1} = 0.211(33)$,
the cyan diamond points in Fig.~\ref{fig:ChPT-fit}
demonstrates this value.
The fit parameters $L_{\pi K}, L_5$, and the scattering lengths
$m_K a_{K \overline{K}}^{I=1}$ at the physical points
(namely, $m_\pi=0.140$ GeV, $m_K=0.494$ GeV)~\cite{Nakamura:2010zzi}
are also summarized in Table~\ref{tb:ChPT-fit},
and the chiral scale $\mu$ is chosen as
the physical $\eta$ mass, namely, $\mu = 0.548$ GeV~\cite{Nakamura:2010zzi}
as it is done in Ref.~\cite{Fu:2011wc}.

\begin{table}[h]
\caption{\label{tb:ChPT-fit}
The fitted $s$-wave scattering lengths $m_K a_K \overline{K}^{I=1}$
at the physical point ($m_\pi=0.140$ GeV, $m_K=0.494$ GeV).
The chiral scale $\mu$ is taken as the physical $\eta$ mass.
}
\begin{tabular*}{\columnwidth}{@{\extracolsep{\fill}}cccc@{}}
\hline
$\chi^2/{\mathrm{dof}}$  & $10^3 \cdot L_{K \overline{K}}^{I=1}$  & $10^3 \cdot L_5$   &
$m_K a_{K \overline{K}}^{I=1}$     \\
\hline
$ 0.656/4 $  & $ -2.41 \pm 1.10$  &   $4.90\pm 4.35 $  & $0.211 \pm 0.033$   \\
\hline
\end{tabular*}
\end{table}

From Fig.~\ref{fig:ChPT-fit}, we can observe that
our lattice simulation results of the scattering lengths
have a large error,
and are in reasonable agreement
with the $\rm SU(3)$ $\chi$PT at NLO.
The fitted values of the $L_5$  and $m_K a_{K \overline{K}}^{I=1}$
have large statistical errors, which reflect the big errors of
our extracted lattice data of $m_K a_{K \overline{K}}^{I=1}$.
Thus we can not claim they have physical meanings.

In this work, we only consider the statistical errors.
The possible sources of the systematic errors on the extrapolated value
of $m_K a_{K \overline{K}}^{I=1}$ mainly contains two parts.
First, according to aforementioned discussions, when extracting
$m_K a_{K \overline{K}}^{I=1}$, we ignore two major systematic errors:
the truncation of the effective range $r$ and
the contributions of the non-Nambu-Goldstone kaons and anti-kaons
in the intermediate states.
These systematic errors of the extracting $m_K a_{K \overline{K}}^{I=1}$
should  propagate through the chiral extrapolation.
Second, the extrapolation to the physical point needs
the experimental value for $m_\pi$,  $m_K$ and $f_\pi$.
The experimental error on these quantities brings
a quite small systematic error
as compared to the corresponding statistical error
and therefore is neglected.

\section{Summary and outlook}
\label{sec:conclude}
In this work we have carried out lattice study of $K \overline{K}$
scattering lengths, and performed a concrete lattice calculation of
$K \overline{K}$ scattering lengths for isospin $I = 1$ channel,
where rectangular diagram plays a vital role,
for the MILC medium coarse ($a=0.15$ fm) lattice ensemble
with the Asqtad improved the staggered dynamical sea quarks.
We employed the technique in Refs.~\cite{Kuramashi:1993ka,Fukugita:1994ve}
(namely, the moving wall sources without gauge fixing)
to reliably calculate  $K \overline{K}$ four-point correlation function,
and observed a clear signal of the attraction for isospin $I=1$ channel,
which is in well accordance with the pioneering lattice studies
on $K \overline{K}$ potential in Refs.~\cite{Mihaly:Ph.D,Mihaly:1999ff}.
Extrapolating the lattice data of
the $s$-wave scattering lengths to the physical point,
we achieved the scattering length
$m_K  a_{K \overline{K}}^{I=1} = 0.211(33)$
directly from lattice simulations.

A good signal can be seen for long time separation range
in the rectangular $R_u$  diagram of $K \overline{K}$ scattering.
We can further reduce the noise by a larger lattice volume
or smaller pion mass,
and hence obtain better results for the scattering length.
Moreover, as pointed above, if we want to obtain the good signals of
the rectangular $R_s$ and vacuum diagrams,
we should choose lattice ensemble with suitable strange quark.

We should understand that the investigation on $K \overline{K}$
scattering at the $I=1$ channel is just first step
in the study of the $K \overline{K}$ scattering.
Since the $K^+K^-$ system is a mixture of the isospin $I=0$ and $1$,
and now there are enough experimental results about its scattering lengths,
the lattice study on $K \overline{K}$ scattering length
at the $I=0$ channel is highly desired.
We are planning a series of lattice studies on it.

\begin{acknowledgements}
This work is supported in part by Fundamental Research Funds
for the Central Universities (2010SCU23002).
We thank MILC Collaboration for using their lattice ensemble.
We should thank Eulogio Oset for his encouraging
and constructive comments, and Zhi-Hui Guo and Hou qing for their kind helps.
The computations for this work were done at AMAX,
CENTOS and HP workstations in the Institute of
Nuclear Science \& Technology, Sichuan University.
\end{acknowledgements}

\appendix
\section{\label{sec:AppendixA}
$I=1$  $K\overline{K}$ scattering amplitudes in terms of quark propagators
}
\begin{eqnarray}
&&\hspace{-0.5cm}\langle O_{K^{+}}(x_4) O_{K^{-}}(x_3)
        O_{K^{+}}^{\dag}(x_2) O_{K^{-}}^{\dag}(x_1) \rangle =\cr
&& {\rm Tr} \left( G^{(u)}(x_{2}, x_{4}) \gamma_{5}
                    G^{(s)}(x_{4}, x_{2}) \gamma_{5} \right) \times \cr
&&     {\rm Tr}\left( G^{(u)}(x_{3}, x_{1}) \gamma_{5}
                    G^{(s)}(x_{1}, x_{3}) \gamma_{5} \right)
\label{appA:term_01} \\
&-& {\rm Tr} \left( G^{(s)}(x_{2}, x_{4}) \gamma_{5}
                    G^{(u)}(x_{4}, x_{3}) \gamma_{5} \times \right. \cr
&& \left.         G^{(s)}(x_{3}, x_{1}) \gamma_{5}
                    G^{(u)}(x_{1}, x_{2}) \gamma_{5} \right)
\label{appA:term_02} \\
&-& {\rm Tr} \left( G^{(u)}(x_{1}, x_{3}) \gamma_{5}
                    G^{(s)}(x_{3}, x_{4}) \gamma_{5} \times \right. \cr
&& \left.                    G^{(u)}(x_{4}, x_{2}) \gamma_{5}
                    G^{(s)}(x_{2}, x_{1}) \gamma_{5} \right)
\label{appA:term_03} \\
&+& {\rm Tr} \left( G^{(u)}(x_{3}, x_{4}) \gamma_{5}
                    G^{(s)}(x_{4}, x_{3}) \gamma_{5} \times \right) \cr
&&     {\rm Tr}\left( G^{(u)}(x_{2}, x_{1}) \gamma_{5}
                    G^{(s)}(x_{1}, x_{2}) \gamma_{5} \right).
\label{appA:term_04}
\end{eqnarray}
\begin{eqnarray}
&&\hspace{-0.5cm}\langle O_{K^{+}}(x_4) O_{K^{-}}(x_3)
        O_{K^{0}}^{\dag}(x_2) O_{\overline{K}^{0}}^{\dag}(x_1)\rangle=\cr
&-&{\rm Tr} \left( G^{(u)}(x_{3}, x_{4}) \gamma_{5}
                    G^{(s)}(x_{4}, x_{2}) \gamma_{5} \right) \times \cr
&&     {\rm Tr}\left( G^{(d)}(x_{2}, x_{1}) \gamma_{5}
                    G^{(s)}(x_{1}, x_{3}) \gamma_{5} \right)
\label{appA:term_05} \\
&-& {\rm Tr} \left( G^{(s)}(x_{3}, x_{4}) \gamma_{5}
                    G^{(u)}(x_{4}, x_{3}) \gamma_{5} \right) \times \cr
&&  {\rm Tr}\left( G^{(s)}(x_{2}, x_{1}) \gamma_{5}
                    G^{(d)}(x_{1}, x_{2}) \gamma_{5} \right).
\label{appA:term_06}
\end{eqnarray}
\begin{eqnarray}
&&\hspace{-0.5cm}
\langle O_{K^{0}}(x_4) O_{\overline{K}^{0}}(x_3)
        O_{K^{+}}^{\dag}(x_2) O_{\overline{K}^{-}}^{\dag}(x_1)\rangle = \cr
&-& {\rm Tr} \left( G^{(d)}(x_{3}, x_{4}) \gamma_{5}
                    G^{(s)}(x_{4}, x_{2}) \gamma_{5} \right) \times \cr
&&     {\rm Tr}\left( G^{(u)}(x_{2}, x_{1}) \gamma_{5}
                    G^{(s)}(x_{1}, x_{3}) \gamma_{5} \right)
\label{appA:term_07} \\
&-& {\rm Tr} \left( G^{(d)}(x_{3}, x_{4}) \gamma_{5}
                    G^{(s)}(x_{4}, x_{3}) \gamma_{5} \right) \times \cr
&&     {\rm Tr}\left( G^{(u)}(x_{2}, x_{1}) \gamma_{5}
                    G^{(s)}(x_{1}, x_{2}) \gamma_{5} \right).
\label{appA:term_08}
\end{eqnarray}
\begin{eqnarray}
&&\hspace{-0.5cm}
\langle O_{K^{0}}(x_4) O_{\overline{K}^{0}}(x_3)
        O_{K^{0}}^{\dag}(x_2) O_{\overline{K}^{0}}^{\dag}(x_1) \rangle = \cr
&& {\rm Tr} \left( G^{(d)}(x_{2}, x_{4}) \gamma_{5}
                    G^{(s)}(x_{4}, x_{2}) \gamma_{5} \right) \times \cr
&&     {\rm Tr}\left( G^{(d)}(x_{3}, x_{1}) \gamma_{5}
                    G^{(s)}(x_{1}, x_{3}) \gamma_{5} \right)
\label{appA:term_09} \\
&-& {\rm Tr} \left( G^{(d)}(x_{2}, x_{4}) \gamma_{5}
                    G^{(s)}(x_{4}, x_{3}) \gamma_{5}  \times \right. \cr
&&\left.            G^{(d)}(x_{3}, x_{1}) \gamma_{5}
                    G^{(s)}(x_{1}, x_{2}) \gamma_{5} \right)
\label{appA:term_10} \\
&-& {\rm Tr} \left( G^{(d)}(x_{3}, x_{4}) \gamma_{5}
                    G^{(s)}(x_{4}, x_{2}) \gamma_{5} \times \right. \cr
&& \left.           G^{(d)}(x_{2}, x_{1}) \gamma_{5}
                    G^{(s)}(x_{1}, x_{2}) \gamma_{5} \right)
\label{appA:term_11} \\
&+& {\rm Tr} \left( G^{(s)}(x_{3}, x_{4}) \gamma_{5}
                    G^{(d)}(x_{4}, x_{3}) \gamma_{5} \right) \times  \cr
&&  {\rm Tr}\left( G^{(s)}(x_{2}, x_{1}) \gamma_{5}
                    G^{(d)}(x_{1}, x_{2}) \gamma_{5} \right).
\label{appA:term_12}
\end{eqnarray}

\section{The analytic expression of $s$-wave $K \overline{K}$ scattering length
for the isospin $I=1$ channel}
\label{sec:AppendixB}
In the isospine limit, both  isospin amplitudes
in $K\overline{K}\rightarrow K\overline{K}$
can be described by two independent amplitudes $T_{ch}$
(the amplitude for the processes  $K^+ K^-\rightarrow K^+ K^-$)
and $T_{neu}$ (that for $\overline{K}^0 K^0\rightarrow K^+ K^-$)~\cite{Oller:1998hw,Guerrero:1998ei,GomezNicola:2001as}.
Z.~H. Guo and J.~A. Oller derived two isospin amplitudes
only in terms of $T_{neu}$~\cite{Guo:2011pa}.
Here we follow the notations and conventions in~\cite{Guo:2011pa},
since the the chiral scale $\mu$ dependence in amplitude is
canceled by the loops and the low energy constants (LECs)~\cite{Guo:2011pa}.

These amplitudes can be decomposed into partial waves $t_l(s)$ according to
$$
T(s,t,u)=16\pi \sum_l (2l+1) t_l(s) P_l(\cos\theta) ,
$$
where $l$ is total angular momentum,
$\theta$ denotes scattering angle in the center-of-mass system.
In the elastic region, the partial wave amplitude $t_l(s)$
can be parameterized by real phase shifts $\delta_l(s)$,
$$
t_l^I(s) = \sqrt{\frac{s}{s-4m_K^2}} \frac{1}{2i} \left\{
e^{2i\delta_l^I(s)}-1 \right\} .
$$
And its real part can be expanded
at threshold (namely, $s=4m_K^2, t=0, u=0$) in terms of the scattering lengths
($a_l^{I=1}$) and effective ranges ($b_l^{I=1}$),
$$
{\rm Re}\,t_l^I(s) = \frac{\sqrt{s}}{2} \, q^{2l}
\Bigl\{ a_l^{I=1} + b_l^{I=1} q^2 + {\cal O} \left(q^4\right) \Bigr\} ,
$$
for the center-of-mass three-momentum of the kaons $q$.
The $s$-wave scattering length is linked to the real part of
the amplitude at threshold by
$$
{\rm Re} \, T^{I=1}_{\rm thr}=
16\pi m_K a_{K\overline{K}}^{I=1} + {\cal O}(q^2) .
$$

The one-loop order analytic expressions of $K \overline{K}$ scattering amplitudes in the isospin limit can be found in Ref.~\cite{Guo:2011pa},
however, no analytic formulae for $s$-wave scattering lengths
were explicitly provided.
We, therefore, will present it for isospin $I=1$ channel.
To this end, we first expand the scattering amplitude at threshold,
namely~\footnote{
Zhi-Hui Guo kindly give the following formulae to me to fitting our lattice data.
In there, we especially thank him.
Without his kind help, we can not finish this work smoothly.
Nevertheless, in this work, we use $f_\pi = 132 MeV$ instead of $93$~MeV
in Guo's original formula.
},
\begin{eqnarray}
T(s,t,u)&=& \frac{2m_K^2}{f_\pi^2} + \frac{32 m_K^2 m_\pi^2}{f_\pi^4}L_5(\mu)
+ \frac{64m_K^4}{f_\pi^4} L_{K \overline{K}}^{I=1}(\mu) \cr
&&- \frac{\left(3m_\eta^2+m_\pi^2\right)^2 + 144m_K^4}{288\pi^2 f_\pi^4}
  - \frac{3 m_K^4}{8 \pi^2 f_\pi^4} \log\frac{m_K^2}{\mu^2} \cr
&&+\frac{1}{432\pi^2 f_\pi^4 (m_\eta^2-m_\pi^2)} \bigg[m_\pi^2(376m_K^4+100m_K^2m_\pi^2 \cr
&&+m_\pi^4 -228m_\eta^2m_K^2
+3m_\eta^2m_\pi^2)\bigg]
\log\frac{m_\pi^2}{\mu^2} \cr
&&+\frac{1}{4320\pi^2 f_\pi^4 (m_\eta^2-m_\pi^2)}
\bigg[15m_\pi^6 -441m_\eta^6 \cr
&& + 3m_\eta^4(716m_K^2+73m_\pi^2) \cr
&&+m_\eta^2(167m_\pi^4-3760m_K^4 -868m_\pi^2m_K^2)\Bigg]
\log\frac{m_\eta^2}{\mu^2} \cr
&& + \overline{J}_{KK}(s=4m_K^2) \frac{4m_K^4}{f_\pi^4} \cr
&& + \overline{J}_{\pi\eta}(s=4m_K^2)
      \frac{\left(3m_\eta^2 -28m_K^2 + m_\pi^2 \right)^2}{54f_\pi^4} ,
\end{eqnarray}
where  $L_{K \overline{K}}^{I=1}(\mu) \equiv
2L_1 + 2L_2 + L_3 - 2L_4  -\frac{1}{2}L_5 + 2L_6 + L_8$
are low-energy constants defined in Ref.~\cite{Gasser:1984gg}
at the chiral symmetry breaking scale $\mu$.
The loop function $\overline{J}_{PQ}$  is
given in Ref.~\cite{Gasser:1984gg}, namely,
\begin{eqnarray}
\overline{J}_{\pi\eta}(s=4m_K^2) &=&
\frac{1}{32\pi^2}\Bigg\{
2+\left(\frac{m_\pi^2 - m_\eta^2}{4m_K^2}-
\frac{m_\pi^2 + m_\eta^2}{m_\pi^2 - m_\eta^2}\right)
\ln \frac{m_\eta^2}{m_\pi^2} \cr
&-& \frac{\nu}{4m_K^2}
\ln\frac{( 4m_K^2 +\nu)^2-(m_\pi^2 - m_\eta^2)^2}
        {( 4m_K^2 -\nu)^2-(m_\pi^2 - m_\eta^2)^2} \Bigg\} ,
\label{jbarrapq}
\end{eqnarray}
with
$\nu = \sqrt{ \left[4m_K^2 - \left(m_\pi + m_\eta\right)^2\right]
\left[4m_K^2 - \left(m_\pi - m_\eta\right)^2\right] }$ .
By plugging the leading order relation
$m_\eta^2 = (4m_K^2-m_\pi^2)/3$~\cite{Guo:2011pa}
we can simplify the above formula as
\begin{eqnarray}
T(s,t,u)&=& \frac{2m_K^2}{f_\pi^2} + \frac{32 m_K^2 m_\pi^2}{f_\pi^4}L_5(\mu)
+ \frac{64m_K^4}{f_\pi^4} L_{K \overline{K}}^{I=1}(\mu) \cr
&&- \frac{5m_K^4}{9\pi^2 f_\pi^4}
  - \frac{3 m_K^4}{8 \pi^2 f_\pi^4} \log\frac{m_K^2}{\mu^2} \cr
&&+\frac{m_\pi^2m_K^2(2m_K^2+5m_\pi^2)}{16\pi^2 f_\pi^4 (m_K^2-m_\pi^2)}
\log\frac{m_\pi^2}{\mu^2} \cr
&&+\frac{9m_\pi^4m_K^2 - 16 m_K^4 m_\pi^2 - 56m_K^6}{144\pi^2 f_\pi^4 (m_K^2-m_\pi^2)}
\log\frac{4m_K^2-m_\pi^2}{3\mu^2} \cr
&& + \overline{J}_{KK}(s=4m_K^2)      \frac{4m_K^4}{f_\pi^4}  \cr
&& + \overline{J}_{\pi\eta}(s=4m_K^2) \frac{32m_K^4}{3f_\pi^4} .
\end{eqnarray}
And the scattering length can be written in a compact form,
\begin{eqnarray}
\label{eq:a1_alg}
a_{K \overline{K}}^{I=1} &=& \frac{m_K}{8\pi f_\pi^2}
\bigg\{ 1 + \frac{16 m_\pi^2}{f_\pi^2}L_5(\mu) +
\frac{32 m_K^2}{f_\pi^2} L_{K \overline{K}}^{I=1}(\mu) \cr
&&+\frac{1}{ (4\pi f_\pi)^2} \chi_{K \overline{K}}^{I=1}(\mu)
\bigg\} ,
\end{eqnarray}
where $\chi_{K \overline{K}}^{I=1}(\mu)$ is the known functions at NLO
which clearly depend on the chiral scale $\mu$ with chiral logarithm terms,
\begin{eqnarray}
\label{eq:chi_kk_1}
\chi_{K \overline{K}}^{I=1}(\mu) &=&
-3m_K^2 \ln \frac{m_K^2}{\mu^2} +
\frac{m_\pi^2}{2}\frac{2m_K^2 + 5m_\pi^2}{m_K^2-m_\pi^2} \ln \frac{m_\pi^2}{\mu^2}\cr
&&+
\frac{9m_\pi^4 - 16 m_K^2 m_\pi^2 - 56m_K^4}{18 (m_K^2-m_\pi^2)}
\log\frac{4m_K^2-m_\pi^2}{3\mu^2}  \cr
&&-\frac{4}{9} m_K^2 + \overline{J}_{\pi\eta}(s=4m_K^2)
\frac{256 \pi^2 m_k^2}{3} .
\end{eqnarray}


\end{document}